\documentclass{article}%
\usepackage{amsfonts}
\usepackage{amsmath}
\usepackage{amssymb}
\usepackage{graphicx}%
\providecommand{\U}[1]{\protect\rule{.1in}{.1in}}
\newtheorem{theorem}{Theorem}
\newtheorem{acknowledgement}[theorem]{Acknowledgement}

\newtheorem{notation}[theorem]{Notation}

\begin{document}

\title{Born--Jordan Quantization and the Equivalence of Matrix and Wave Mechanics}
\author{Maurice A. de Gosson\thanks{maurice.de.gosson@univie.ac.at,
maurice.degosson@gmail.com}\\University of Vienna\\Faculty of Mathematics, NuHAG}
\maketitle

\begin{abstract}
The aim of the famous Born and Jordan 1925 paper was to put Heisenberg's
matrix mechanics on a firm mathematical basis. Born and Jordan showed that if
one wants to ensure energy conservation in Heisenberg's theory it is necessary
and sufficient to quantize observables following a certain ordering rule. One
apparently unnoticed consequence of this fact is that Schr\"{o}dinger's wave
mechanics cannot be equivalent to Heisenberg's more physically motivated
matrix mechanics unless its observables are quantized using \textit{this}
rule, and \textit{not} the more symmetric prescription proposed by Weyl in
1926, which has become the standard procedure in quantum mechanics. This
observation confirms the superiority of Born-Jordan quantization, as already
suggested by Kauffmann. We also show how to explicitly determine the
Born--Jordan quantization of arbitrary classical variables, and discuss the
conceptual advantages in using this quantization scheme. We finally suggest
that it might be possible to determine the correct quantization scheme by
using the results of weak measurement experiments.

\end{abstract}

\section{INTRODUCTION}

In the Schr\"{o}dinger picture of quantum mechanics (wave mechanics), the
operators are constant (unless they are explicitly time-dependent), and the
states evolve in time: $|\psi(t)\rangle=U(t,t_{0})|\psi(t_{0})\rangle$ where
\begin{equation}
U(t,t_{0})=e^{iH_{\mathcal{S}}(t-t_{0})/\hslash} \label{unit}%
\end{equation}
is a family of unitary operators; the time evolution of $|\psi\rangle$ is thus
governed by Schr\"{o}dinger's equation
\begin{equation}
i\hbar\frac{d\psi}{dt}=H_{\mathcal{S}}\psi; \label{sch}%
\end{equation}
$H_{\mathcal{S}}$ is an operator associated with the classical Hamiltonian
function $H$ by some \textquotedblleft quantization rule\textquotedblright. In
the Heisenberg picture (matrix mechanics), the state vectors are
time-independent operators that incorporate a dependency on time, while an
observable $A_{\mathcal{S}}$ in the Schr\"{o}dinger picture becomes a
time-dependent operator $A_{\mathcal{H}}(t)$ in the Heisenberg picture; this
time dependence satisfies the Heisenberg equation%
\begin{equation}
i\hbar\frac{dA_{\mathcal{H}}}{dt}=i\hbar\frac{\partial A_{\mathcal{H}}%
}{\partial t}+[A_{\mathcal{H}},H_{\mathcal{H}}]. \label{hei}%
\end{equation}

Schr\"{o}dinger \cite{schr} (and, independently, Eckart \cite{eck}) attempted
to\ prove shortly after the publication of Heisenberg's result that wave
mechanics and matrix mechanics were mathematically equivalent. Both proofs
contained flaws, and one had to wait until von Neumann's \cite{von} seminal
work for a rigorous proof of the equivalence of both theories (see the
discussions in Madrid Casado \cite{madrid} and Muller \cite{muller}; both
papers contain a wealth of historical details; also see van der Waerden's
\cite{vanw} very interesting discussion of an unpublished letter of Pauli
about the (non)equivalence of wave mechanics and matrix mechanics). We will
not bother with the technical shortcomings of Schr\"{o}dinger's and Eckart's
approaches here, but rather focus on one, perhaps more fundamental, aspect
which seems to have been overlooked in the literature. We observe that it is
possible to go from the Heisenberg picture to the Schr\"{o}dinger picture (and
back) using the following simple argument (see for instance Messiah
\cite{Messiah} or Schiff \cite{Schiff}): a ket
\begin{equation}
|\psi_{\mathcal{S}}(t)\rangle=U(t,t_{0})|\psi_{\mathcal{S}}(t_{0}%
)\rangle\label{psis}%
\end{equation}
in the Schr\"{o}dinger picture becomes, in the Heisenberg picture, the
constant ket
\begin{equation}
|\psi_{\mathcal{H}}\rangle=U(t,t_{0})^{\ast}|\psi_{\mathcal{S}}(t)\rangle
=|\psi_{\mathcal{S}}(t_{0})\rangle\label{psih}%
\end{equation}
whereas an observable $A_{\mathcal{S}}$ becomes
\begin{equation}
A_{\mathcal{H}}(t)=U(t,t_{0})^{\ast}A_{\mathcal{S}}U(t,t_{0}); \label{ah}%
\end{equation}
in particular the Hamiltonian is%
\begin{equation}
H_{\mathcal{H}}(t)=U(t,t_{0})^{\ast}H_{\mathcal{S}}U(t,t_{0}). \label{hh}%
\end{equation}
Taking $t=t_{0}$ this relation implies that $H_{\mathcal{H}}(t_{0}%
)=H_{\mathcal{S}}$; now in the Heisenberg picture energy is constant, so the
Hamiltonian operator $H_{\mathcal{H}}(t)$ must be a constant of the motion. It
follows that $H_{\mathcal{H}}(t)=H_{\mathcal{S}}$ for all times $t$ and hence
both operators $H_{\mathcal{H}}$ and $H_{\mathcal{S}}$ must be quantized using
the \emph{same} rules. A consequence of this property is that if we believe
that Heisenberg's \textquotedblleft matrix mechanics\textquotedblright\ is
correct and is equivalent to Schr\"{o}dinger's theory, then the Hamiltonian
operator appearing in the Schr\"{o}dinger equation (\ref{sch}) \emph{must} be
quantized using the Born--Jordan rule, and not, as is usual in quantum
mechanics, the Weyl quantization rule.

\begin{notation}
Real position and momentum variables are denoted $q,p$; more generally, for
systems with $n$ degrees of freedom we write $q=(q_{1},..,q_{n})$,
$p=(p_{1},...,p_{n})$. The boldface letters $\mathbf{q},\mathbf{p}$ are used
to denote the corresponding quantum observables. Similarly, the quantum
operator associated with a classical observable $A$ is denoted by $\mathbf{A}$
and we write $A\longleftrightarrow\mathbf{A}$. It is assumed throughout that
this correspondence (\textquotedblleft quantization\textquotedblright) is
\emph{linear}.
\end{notation}

\section{THE\ BORN AND\ JORDAN\ ARGUMENT}

We begin by shortly exposing the main arguments in Born and Jordan's paper
\cite{bj}.

The paper of Born and Jordan was an attempt to put Heisenberg's
\textquotedblleft magical paper\textquotedblright\ \cite{hei} on a firm basis
(see Aitchison \textit{et al.} \cite{ai}) and van der Waerden \cite{vanw} for
interesting discussions of Heisenberg's paper from a modern\ point of view).
Following Heisenberg's paper \cite{hei} Born and Jordan considered in
\cite{bj} square infinite matrices%
\begin{equation}
\mathbf{a}=(a(n,m))=%
\begin{pmatrix}
a(00) & a(01) & a(02) & \cdot\cdot\cdot\\
a(10) & a(11) & a(12) & \cdot\cdot\cdot\\
a(20) & a(21) & a(22) & \cdot\cdot\cdot\\
\cdot\cdot\cdot & \cdot\cdot\cdot & \cdot\cdot\cdot & \cdot\cdot\cdot
\end{pmatrix}
\label{1}%
\end{equation}
where the $a(nm)$ are what they call \textquotedblleft ordinary
quantities\textquotedblright, \textit{i.e.} scalars; we will call these
infinite matrices (for which we always use boldface letters)
\emph{observables}. In particular Born and Jordan introduce momentum and
position observables $\mathbf{p}$ and $\mathbf{q}$ and matrix functions
$\mathbf{H}(\mathbf{p,q)}$ of these observables, which they call
\textquotedblleft Hamiltonians\textquotedblright. Following Heisenberg, they
assume that the equations of motion for $\mathbf{p}$ and $\mathbf{q}$ are
formally the same as in classical theory, namely%
\begin{equation}
\mathbf{\dot{q}}=\frac{\partial\mathbf{H}}{\partial\mathbf{p}}\text{ \ ,
\ }\mathbf{\dot{p}}=-\frac{\partial\mathbf{H}}{\partial\mathbf{q}}; \label{3}%
\end{equation}
limiting themselves deliberately to Hamiltonians which are polynomials in the
observables $\mathbf{p}$, $\mathbf{q}$, that is linear combinations of
monomials which are products of terms
\begin{equation}
\mathbf{H=p}^{s}\mathbf{q}^{r} \label{4}%
\end{equation}
they define the derivatives in (\ref{3}) by the formulas, and show that the
observables $\mathbf{p}$ and $\mathbf{q}$ satisfy the commutation relation%
\begin{equation}
\mathbf{pq}-\mathbf{qp}=-i\hbar\mathbf{1} \label{2}%
\end{equation}
where $\mathbf{1}$ is the identity matrix; from this follows the more general
identity%
\begin{equation}
\mathbf{p}^{m}\mathbf{q}^{n}-\mathbf{q}^{n}\mathbf{p}^{m}=-i\hbar m\sum
_{\ell=0}^{n-1}\mathbf{q}^{n-1-\ell}\mathbf{p}^{m-1}\mathbf{q}^{\ell}.
\label{7}%
\end{equation}
Born and Jordan next proceed to derive the fundamental laws of quantum
mechanics. In particular, pursuing their analogy with classical mechanics,
they want to prove that energy is conserved; identifying the values of the
Hamiltonian $\mathbf{H}$ with the energy of the system, they impose the
condition $\mathbf{\dot{H}=0}$ and show that this condition requires that%
\begin{gather}
-i\hbar\mathbf{\dot{q}}=\mathbf{Hq-qH}\text{ }\label{10}\\
-i\hbar\text{\ }\mathbf{\dot{p}}=\mathbf{Hp-pH.}%
\end{gather}
Comparing with the Hamilton-like equations (\ref{3})\ this condition is in
turn equivalent to
\begin{gather}
\mathbf{Hq-qH}=-i\hbar\frac{\partial\mathbf{H}}{\partial\mathbf{p}}%
\label{11}\\
\text{ }\mathbf{Hp-pH}=i\hbar\frac{\partial\mathbf{H}}{\partial\mathbf{q}}.
\end{gather}
Now comes the crucial step. Given a classical Hamiltonian $H(p,q)=p^{s}q^{r}$
they ask how one should choose the observable $\mathbf{H}(\mathbf{p}%
,\mathbf{q})$ so that these identities hold. Using the commutation formula
(\ref{7}) Born and Jordan show that the \emph{only} possible choice is%
\begin{equation}
\mathbf{H}(\mathbf{p},\mathbf{q})=\frac{1}{s+1}\sum_{\ell=0}^{s}%
\mathbf{p}^{s-\ell}\mathbf{q}^{r}\mathbf{p}^{\ell}; \label{12}%
\end{equation}

\section{BORN-JORDAN\ QUANTIZATION}

Born and Jordan thus proved --rigorously-- that the only way to quantize
polynomials in a way consistent with Heisenberg's ideas was to use the rule%
\begin{equation}
p^{s}q^{r}\overset{\mathrm{BJ}}{\longrightarrow}\frac{1}{s+1}\sum_{\ell=0}%
^{s}\mathbf{p}^{s-\ell}\mathbf{q}^{r}\mathbf{p}^{\ell};\label{bj1}%
\end{equation}
equivalently, using the commutation relations (\ref{7}):%
\begin{equation}
p^{s}q^{r}\overset{\mathrm{BJ}}{\longrightarrow}\frac{1}{r+1}\sum_{j=0}%
^{r}\mathbf{q}^{r-j}\mathbf{p}^{s}\mathbf{q}^{j}.\label{bj2}%
\end{equation}
In their subsequent publication \cite{bjh} with Heisenberg they show that
their constructions extend \textit{mutatis mutandis} to systems with an
arbitrary number of degrees of freedom. We will call this rule (and its
extension to higher dimensions) the \emph{Born--Jordan }(BJ)\emph{
quantization rule}. Weyl \cite{Weyl} proposed, independently, some time later
(1926) another rule leading to the replacement of (\ref{bj1}) with
\begin{equation}
p^{s}q^{r}\overset{\mathrm{Weyl}}{\longrightarrow}\frac{1}{2^{s}}\sum_{\ell
=0}^{s}%
\begin{pmatrix}
s\\
\ell
\end{pmatrix}
\mathbf{p}^{s-\ell}\mathbf{q}^{r}\mathbf{p}^{\ell}.\label{w2}%
\end{equation}
Both quantizations are thus not equivalent; one can show (Turunen
\cite{Turunen}) by induction that the Weyl and Born--Jordan rules differ for
all polynomials $p^{s}q^{r}$ with $s\geq2$ and $r\geq2$. (We would like to
take the opportunity to express our thanks to Maciej Blaszak for having
pointed out an error in the first version of this paper). As Kauffmann
\cite{Kauffmann} observes, Weyl's rule is the single most symmetrical operator
ordering, whereas the BJ quantization is the equally weighted average of all
the operator orderings (also see Song \cite{Song}).

These facts have the following consequence: if we insist that the Heisenberg
and Schr\"{o}dinger pictures be \emph{equivalent}, then we \emph{must}
quantize the Hamiltonian in Schr\"{o}dinger's equation using BJ quantization.
In fact, recall from formula (\ref{hh}) that the Heisenberg and
Schr\"{o}dinger Hamiltonians are related by
\[
H_{\mathcal{H}}(t)=U(t,t_{0})^{\ast}H_{\mathcal{S}}U(t,t_{0}).
\]
Since $H_{\mathcal{H}}(t)$ is a constant of the motion we have $H_{\mathcal{H}%
}(t)=H_{\mathcal{H}}(t_{0})$ and hence $H_{\mathcal{H}}(t)=H_{\mathcal{S}}$ so
the Heisenberg and Schr\"{o}dinger Hamiltonians $H_{\mathcal{H}}(t)$ and
$H_{\mathcal{H}}$ must be identical. But the condition $H_{\mathcal{H}%
}(t)=H_{\mathcal{H}}(t_{0})=H_{\mathcal{H}}$ means that $H_{\mathcal{H}}$ and
hence $H_{\mathcal{S}}$ must be quantized using the Born and Jordan prescription.

An obvious consequence of these considerations is that if one uses in the
Schr\"{o}dinger picture the Weyl quantization rule (or any other quantization
rule), we obtain two different renderings of quantum mechanics. This
observation seems to be confirmed by Kauffmann's \cite{Kauffmann} interesting
discussion of the non-physicality of Weyl quantization.

We have been considering the quantization of polynomials for simplicity; in de
Gosson and Luef \cite{GoLuBJ} and de Gosson \cite{TRANSAM} we have shown in
detail how to Born--Jordan quantize arbitrary functions of the position and
momentum variables.

\section{GENERALIZATION\ TO\ ARBITRARY\ OBSERVABLES}

The Weyl quantization rule (\ref{w2}) can be viewed as a particular case of a
very general rule, which we call the \textquotedblleft$\tau$%
-rule\textquotedblright. Let us first consider a very simple example, that of
the monomial $p^{2}q$, for which we gave in (\ref{bjw}) both the BJ and the
Weyl quantizations. Let now $\tau$ be an arbitrary real number and consider
the following modification of the Weyl quantization rule (\ref{bjw})%
\begin{equation}
p^{2}q\overset{\tau}{\longrightarrow}(1-\tau)^{2}\mathbf{p}^{2}\mathbf{q}%
+2(1-\tau)\tau\mathbf{pqp}+\tau^{2}\mathbf{qp}^{2}\label{bjw2}%
\end{equation}
(it reduces to the latter if we choose $\tau=\frac{1}{2}$). If we integrate
the right-hand side from $0$ to $1$ in $\tau$ we get%
\[
\int_{0}^{1}\left[  (1-\tau)^{2}\mathbf{p}^{2}\mathbf{q}+2(1-\tau
)\tau\mathbf{pqp}+\tau^{2}\mathbf{qp}^{2}\right]  d\tau=\frac{1}{3}%
(\mathbf{p}^{2}\mathbf{q}+\mathbf{pqp}+\mathbf{qp}^{2})
\]
which is precisely the BJ quantization of the monomial $p^{2}q$ (it however
coincides with the Weyl quantization: see below). More generally, we define
the \textquotedblleft$\tau$-quantization rule\textquotedblright\ for monomials
by
\begin{equation}
p^{s}q^{r}\overset{\tau}{\longrightarrow}\frac{1}{2^{s}}\sum_{\ell=0}^{s}%
\begin{pmatrix}
s\\
\ell
\end{pmatrix}
(1-\tau)^{\ell}\tau^{s-\ell}\mathbf{p}^{s-\ell}\mathbf{q}^{r}\mathbf{p}^{\ell
}.\label{tauq1}%
\end{equation}
Again it reduces to the Weyl quantization when $\tau=\frac{1}{2}$; if we
integrate the right-hand side from $0$ to $1$ in $\tau$ and observing that it
follows from the properties of the beta function that%
\[
\int_{0}^{1}(1-\tau)^{\ell}\tau^{s-\ell}d\tau=B(\ell+1,s-\ell)=\frac
{\ell!(s-\ell)!}{(s+1)!}%
\]
we recover the BJ quantization rule (\ref{bj1}). This essential observation
allows us to define the BJ quantization of an arbitrary classical observable.
While we have done this from an operator-theoretical point of view in de
Gosson \cite{TRANSAM} and de Gosson and Luef \cite{GoLuBJ}, we will follow
here a more physical approach, along the lines of Kauffmann \cite{Kauffmann}
with some modifications. We are working in $n$-dimensional configuration
space, since it does not add any difficulty. The Weyl quantization
$\mathbf{A}_{\mathrm{W}}(\mathbf{q},\mathbf{p})$ of a general observable
$A(q,p)$ is unambiguously defined in its configuration representation by the
Fourier transform
\begin{equation}
\langle q_{2}|\mathbf{A}_{\mathrm{W}}|q_{1}\rangle=\left(  \tfrac{1}{2\pi
\hbar}\right)  ^{n}\int e^{\frac{i}{\hbar}p(q_{2}-q_{1})}A(\tfrac{1}{2}%
(q_{1}+q_{2}),p)d^{n}p\mathbf{.}\label{weyl1}%
\end{equation}
Define similarly $\tau$-quantization $A\overset{\tau}{\longrightarrow
}\mathbf{A}_{\tau}$ in the configuration representation by
\begin{equation}
\langle q_{2}|\mathbf{A}_{\tau}|q_{1}\rangle=\left(  \tfrac{1}{2\pi\hbar
}\right)  ^{n}\int e^{\frac{i}{\hbar}p(q_{2}-q_{1})}A(\tau q_{1}+(1-\tau
)q_{2}),p)d^{n}p\mathbf{;}\label{tauq2}%
\end{equation}
of course $\mathbf{A}_{1/2}=\mathbf{A}_{\mathrm{W}}$. The BJ quantization
$\mathbf{A}_{\mathrm{BJ}}$ is then defined as being the average of all the
$\tau$-quantizations of $A(q,p)$ when the parameter $\tau$ goes from $0$ to
$1$:
\begin{equation}
\mathbf{A}_{\mathrm{BJ}}=\int_{0}^{1}\mathbf{A}_{\tau}d\tau;\label{abjdef}%
\end{equation}
it follows from formula (\ref{tauq2}) that $\mathbf{A}_{\mathrm{BJ}}$ has the
following configuration representation:%
\begin{gather}
\langle q_{2}|\mathbf{A}_{\mathrm{BJ}}|q_{1}\rangle=\left(  \tfrac{1}%
{2\pi\hbar}\right)  ^{n}\int e^{\frac{i}{\hbar}p(q_{2}-q_{1})}\widetilde{A}%
(q_{1},q_{2},p)d^{n}p\label{ab1}\\
\text{with \ }\widetilde{A}(q_{1},q_{2},p)=\int_{0}^{1}A(\tau q_{1}%
+(1-\tau)q_{2}),p)d\tau.\label{ab2}%
\end{gather}

The correspondence $A\overset{\mathrm{BJ}}{\longrightarrow}\mathbf{A}%
_{\mathrm{BJ}}$ thus defined reduces to the correspondence (\ref{bj1}),
(\ref{bj2}) in the monomial case; it moreover has the property (shared with
Weyl quantization) that to a real classical observable $A$ it associates a
self-adjoint operator $\mathbf{A}_{\mathrm{BJ}}$ (this property, which is
essential for any honest quantization theory, is not satisfied by the
\textquotedblleft$\tau$-quantization rule\textquotedblright\ $A\overset{\tau
}{\longrightarrow}\mathbf{A}_{\tau}$, which is hence unphysical). It is easy
to show using the formulas above that the BJ and Weyl quantization of
Hamiltonians of the usual type \textquotedblleft kinetic energy plus
potential\textquotedblright\ are the same; we have shown in \cite{TRANSAM} and
\cite{GoLuBJ} that BJ and Weyl quantization coincide for all Hamiltonians of
the type%
\begin{equation}
H=\sum_{j=1}^{n}\frac{1}{2m_{j}}(p_{j}-A_{j}(q,t))^{2}+U(q,t) \label{hmax}%
\end{equation}
where the vector and scalar potentials $A_{j}$ and $U$ depend on
$q=(q_{1},...,q_{n})$ (and possibly on time $t$); this quantization is given
by the usual formula%
\begin{equation}
\mathbf{H}=\sum_{j=1}^{n}\frac{1}{2m_{j}}\left(  -i\hbar\frac{\partial
}{\partial x_{j}}-A_{j}(q,t)\right)  ^{2}+U(q,t) \label{hmaxq}%
\end{equation}
(Messiah \cite{Messiah}, Schiff \cite{Schiff}).

Let us briefly discuss in this context the property of canonical covariance.
This property singles out Weyl quantization among all possible quantizations;
it is probably thanks to this peculiarity that Weyl quantization superseded
(at least among mathematical physicists) the BJ (and other possible
quantization schemes). It is a very strong property (see the discussion at the
end of the paper); it has allowed us to prove in \cite{gohi} that Hamiltonian
mechanics and quantum mechanics (when quantized using Weyl's rule) are
mathematically equivalent theories, i.e. that one can derive Schr\"{o}dinger's
equation from Hamilton's equations of motion, and vice versa. Canonical
covariance means the following: let $\operatorname*{Sp}(n)$ be the symplectic
group of the $n$-dimensional configuration space; it consists of all linear
canonical transformations of the corresponding $2n$-dimensional phase space
(we have given an elementary construction of $\operatorname*{Sp}(n)$ in de
Gosson \cite{egg}). The elements of $\operatorname*{Sp}(n)$ are identified
with $2n\times2n$ matrices $S$ (\textquotedblleft symplectic
matrices\textquotedblright) satisfying the condition $S^{T}JS=J$ where the
superscript $T$ indicates transposition and $J=%
\begin{pmatrix}
0 & I\\
-I & 0
\end{pmatrix}
$ where $0$ and $I$ are the zero and identity $n\times n$ matrices. Now, to
every symplectic matrix $S$ one can associate two unitary operators
$\pm\widehat{S}$ acting on $L^{2}(\mathbb{R}^{n})$ (the square integrable
functions); the set of all these operators form a group, the\textit{
metaplectic group} $\operatorname*{Mp}(n)$ (see de Gosson \cite{Birk1} for a
detailed study of that group). The property of canonical covariance for a
quantization rule $A\longleftrightarrow\mathbf{A}$ means that for every
symplectic matrix $S$ we must have $A\circ S\longleftrightarrow\widehat{S}%
\mathbf{A}\widehat{S}^{-1}$ ($A\circ S$ is the new observable $A\circ
S(q,p)=A(S(q,p))$). (Thus, to a symplectic transformation of the coordinates
in a classical observable corresponds at the operator level to conjugation by
the corresponding metaplectic operator). It is now a mathematical theorem that
there is only one quantization rule which enjoys this property, and this is
Weyl quantization. Therefore, if we use BJ quantization in place of Weyl
quantization, we will loose canonical covariance for all observables who are
not quantized the \textquotedblleft Weyl way\textquotedblright. But this
observation has no drastic consequences, because as we just mentioned the Weyl
and BJ quantizations of all physical Hamiltonians (\ref{hmax}) are the same,
and will thus have the property of canonical covariance. And there is another
case where this remains true: formula (\ref{w2}) implies that monomials
$q_{j}^{2}$, $p_{j}^{2}$, $p_{j}q_{j}$ (and, of course $p_{j}q_{k})$ have the
same quantization in both schemes; it easily follows that the same true for
the generalized harmonic oscillator%
\begin{equation}
H(p,q)=\sum_{j,k=1}^{n}a_{j}p_{j}^{2}+2b_{j}p_{j}q_{k}+c_{j}q_{j}^{2}.
\label{gho}%
\end{equation}

\section{DISCUSSION}

One might wonder at this point whether it is possible to distinguish both
quantizations at all. It follows from the discussion above that as far as
ordinary Hamiltonians (\ref{hmax}) or generalized oscillators (\ref{gho}) are
considered, we cannot. However there is a conceptually very important reason
for which BJ quantization should be taken very seriously; it is related to the
issue of \textit{dequantization} (or \textquotedblleft
classicization\textquotedblright). Besides being canonically covariant, the
Weyl rule has a very important, but rather unwelcome, property: it is
one-to-one invertible because \emph{every} continuous operator can be written
uniquely as a Weyl operator (for a mathematical proof see \textit{e.g.} de
Gosson \cite{Birk1,Birk2}). This invertibility means that every quantum
observable has a (unique) classical counterpart, and this is physically not
tenable. The situation is very different when one uses BJ quantization. Let us
explain this in some detail. We begin with the following observation, which is
simple and subtle at the same time. Consider the BJ quantization
$\mathbf{A}_{\mathrm{BJ}}\overset{\mathrm{BJ}}{\longleftrightarrow}A$ of some
classical observable $A$. Born--Jordan operators are continuous operators,
hence we can also view $\mathbf{A}_{\mathrm{BJ}}$ as a Weyl operator:
$\mathbf{A}_{\mathrm{BJ}}=\mathbf{B}_{\mathrm{W}}\overset{\mathrm{W}%
}{\longleftrightarrow}B$ where $B$ is generally different from $A$. In de
Gosson \cite{TRANSAM} and de Gosson and Luef \cite{GoLuBJ} we have proven that
the phase space Fourier transforms $\mathcal{F}A$ and $\mathcal{F}B$ of the
classical observables $A$ and $B$ are related by the formula%
\begin{equation}
\mathcal{F}B(q,p)=\left(  \frac{2\hbar}{pq}\sin\frac{pq}{2\hbar}\right)
\mathcal{F}A(q,p). \label{fafb}%
\end{equation}
This formula implies that BJ quantization is neither one-to-one, nor
invertible. In fact, every operator $\mathbf{A}$ has infinitely many
precursors $A\overset{\mathrm{BJ}}{\longleftrightarrow}\mathbf{A}%
_{\mathrm{BJ}}$. Since quantization is linear, it is sufficient to verify this
statement for $\mathbf{A}_{\mathrm{BJ}}=0$; writing again $\mathbf{A}%
_{\mathrm{BJ}}=\mathbf{B}_{\mathrm{W}}\overset{\mathrm{W}}{\longleftrightarrow
}B$ we must have $B=0$ (and hence $\mathcal{F}B=0$) since the Weyl
correspondence is one-to-one. In view of formula (\ref{fafb}) this implies
that $A$ is any observable such that%
\[
\left(  \frac{2\hbar}{pq}\sin\frac{pq}{2\hbar}\right)  \mathcal{F}A(q,p)=0.
\]
Since the factor in front of $\mathcal{F}A(q,p)$ vanishes for all $(q,p)$ such
that $pq=2N\pi\hbar$ ($N$ an integer $\neq0$), this equality will hold for any
classical observable whose Fourier transform vanishes outside the sets of
phase space defined by these conditions; there are of course infinitely such
choices. (See the interesting potential consequences for the limit
$\hbar\rightarrow0$ discussed by Kauffmann \cite{Kauffmann}). Similar
considerations show that the correspondence $A\overset{\mathrm{BJ}%
}{\longleftrightarrow}\mathbf{A}_{\mathrm{BJ}}$ is in general not invertible.
In fact,\ if it were, we could write each Weyl operator $\mathbf{B}%
_{\mathrm{W}}$ would correspond a Born--Jordan operator $\mathbf{A}%
_{\mathrm{BJ}}$ such that $\mathbf{A}_{\mathrm{BJ}}=\mathbf{B}_{\mathrm{W}}$.
The corresponding classical observable $A$ would then be determined by
(\ref{fafb}), but this is generally not possible because of the zeroes of the
sine term.

It would certainly be interesting and useful to have explicit examples; the
calculations are rather technical, and part of work in progress \cite{gotu}.

To conclude, if we believe in the equivalence of Heisenberg's matrix mechanics
and Schr\"{o}dinger's wave mechanics, then we must quantize both theories
using the same correspondence. Matrix mechanics seems to be more physically
motivated, being based on a natural notion, that of conservation of energy,
which leads mathematically to the BJ quantization scheme, while there is no
reason in Schr\"{o}dinger's theory to choose one particular quantization. This
provides strong evidence that Born--Jordan quantization might very well be the
right choice in quantum mechanics. Of course, to sustain this conjecture, it
would be of primordial importance to test it experimentally. We suggest this
could be done using weak measurements: as we have shown in \cite{goab}, the
notion of weak value can be expressed in two different ways, yielding
different numerical results, depending on whether one uses Weyl or BJ
quantization. Suppose in fact we have a classical observable $A$; we denote by
$\mathbf{A}_{\mathrm{W}}$ and $\mathbf{A}_{\mathrm{BJ}}$ the corresponding
Weyl and BJ quantizations. Let $|\psi\rangle$ be a pre-selected state and
$|\phi\rangle$ a post-selected state; if these states are non-orthogonal the
weak values of $\mathbf{A}_{\mathrm{W}}$ and $\mathbf{A}_{\mathrm{BJ}}$ with
respect to the pair $(\phi,\psi)$ are the complex numbers%
\[
\langle\mathbf{A}_{\mathrm{W}}\rangle_{\text{weak}}^{\phi,\psi}=\frac
{\langle\phi|\mathbf{A}_{\mathrm{W}}|\psi\rangle}{\langle\phi|\psi\rangle
}\text{ \ , \ }\langle\mathbf{A}_{\mathrm{BJ}}\rangle_{\text{weak}}^{\phi
,\psi}=\frac{\langle\phi|\mathbf{A}_{\mathrm{BJ}}|\psi\rangle}{\langle
\phi|\psi\rangle}.
\]
We have shown that $\langle\mathbf{A}_{\mathrm{W}}\rangle_{\text{weak}}%
^{\phi,\psi}$ can be calculated by averaging $A$ over the complex phase space
function%
\[
\rho^{\phi,\psi}(q,p)=\frac{W(\phi,\psi)(q,p)}{\langle\phi|\psi\rangle}%
\]
where%
\[
W(\phi,\psi)(q,p)=\left(  \tfrac{1}{2\pi\hbar}\right)  ^{n}\int e^{-\frac
{i}{\hbar}pq^{\prime}}\phi(q+\tfrac{1}{2}q^{\prime})\psi^{\ast}(q-\tfrac{1}%
{2}q^{\prime})d^{n}q^{\prime}%
\]
is the cross-Wigner transform. On the other hand $\langle\mathbf{A}%
_{\mathrm{W}}\rangle_{\text{weak}}^{\phi,\psi}$ is obtained similarly, but by
averaging $A$ this time over%
\[
\rho_{\mathrm{BJ}}^{\phi,\psi}(q,p)=\frac{W_{\mathrm{BJ}}(\phi,\psi
)(q,p)}{\langle\phi|\psi\rangle}%
\]
where $W_{\mathrm{BJ}}(\phi,\psi)$ is the modified cross-Wigner transform
defined by%
\[
W_{\mathrm{BJ}}(\phi,\psi)=W(\phi,\psi)\ast\mathcal{F}\Theta-
\]

\begin{acknowledgement}
This work has been supported by a grant from the Austrian Research Agency FWF
(Projektnummer P20442-N13). I take the opportunity to extend my warmest thanks
to Glen Dennis, who was kind enough to point out errors and misprints.%
\[
\]

\end{acknowledgement}


\begin{thebibliography}{99}                                                                                               %


\bibitem {ai}I. J. R. Aitchison, D. A. MacManus, and T. M. Snyder,
Understanding Heisenberg's \textquotedblleft magical\textquotedblright\ paper
of July 1925: A new look at the calculational details, Am. J. Phys\textit{.}
\textbf{72}, 1370--1379 (2004).

\bibitem {bj}M. Born, P. Jordan, Zur Quantenmechanik, Z. Physik \textbf{34},
858--888 (1925); English translation in: M. Jammer \textit{The Conceptual
Development of Quantum Mechanics}. New York: McGraw-Hill (1966); 2nd ed: New
York: American Institute of Physics (1989).

\bibitem {bjh}M. Born, W. Heisenberg, and P. Jordan, Zur Quantenmechanik II,
Z. Physik \textbf{35}, 557--615 (1925); English translation in: M. Jammer
\textit{The Conceptual Development of Quantum Mechanics}. New York:
McGraw-Hill (1966); 2nd ed: New York: American Institute of Physics (1989).

\bibitem {eck}C. Eckart, Operator Calculus and the Solution of the Equation of
Quantum Dynamics, Phys. Rev. \textbf{28}, 711--726 (1926).

\bibitem {Birk1}M. de Gosson,\textit{ Symplectic Geometry and Quantum
Mechanics.} Birkh\"{a}user, Basel, (2006).

\bibitem {Birk2}M. de Gosson, \textit{Symplectic Methods in Harmonic Analysis
and in Mathematical Physics}. Birkh\"{a}user, Basel (2011).

\bibitem {TRANSAM}M. de Gosson, Symplectic covariance properties for Shubin
and Born-Jordan pseudo-differential operators. Trans. Amer. Math.
Soc\textit{.} \textbf{365}(6), 3287--3307 (2013).

\bibitem {mdgiop}M. de Gosson, Born--Jordan Quantization and the Uncertainty
Principle. J. Phys. A: Math. Theor. \textbf{46} 445301 (2013).

\bibitem {egg}M. de Gosson, The symplectic egg in quantum and classical
mechanics. Amer. J. of Phys. \textbf{81}(5) (2013).

\bibitem {goab}M. de Gosson, L. D. Abreu, Weak values and Born-Jordan
quantization. Quantum Theory: Reconsiderations of Foundations 6. Vol. 1508.
No. 1. AIP Publishing, (2012).

\bibitem {gohi}M. de Gosson, B. Hiley, Imprints of the Quantum World in
Classical Mechanics. \textit{Found. Phys}. \textbf{41}(9) (2011).

\bibitem {GoLuBJ}M. de Gosson, F. Luef, Preferred quantization rules:
Born-Jordan versus Weyl. The pseudo-differential point of view, J.
Pseudo-Differ. Oper. Appl. \textbf{2}(1), 115--139 (2011).

\bibitem {gotu}M. de Gosson, V. Turunen, Dequantization and the Born--Jordan
Correspondence, Preprint 2014.

\bibitem {hei}W. Heisenberg, \"{U}ber quantentheoretische Umdeutung
kinematischer und mechanischer Beziehungen, Z. Physik \textbf{33}, 879--893 (1925).

\bibitem {Jammer}M. Jammmer, \textit{The Conceptual Development of Quantum
Mechanics}. New York: McGraw-Hill (1966); 2nd ed: New York: American Institute
of Physics (1989).

\bibitem {Kauffmann}S. K. Kauffmann, Unambiguous Quantization from the Maximum
Classical Correspondence that Is Self-consistent: The Slightly Stronger
Canonical Commutation Rule Dirac Missed, Found. Phys. \textbf{41}(5), 805--819 (2011).

\bibitem {Littlejohn}R. G. Littlejohn, The semiclassical evolution of wave
packets, Phys. Rep. \textbf{138}(4-5), 193--291 (1986).

\bibitem {madrid}C. M. Madrid Casado, A brief history of the mathematical
equivalence between the two quantum mechanics, Latin American Journal of
Physics Education \textbf{2}(2), 104--108 (2008).

\bibitem {Messiah}A. Messiah, \textit{Quantum Mechanics} (Vol. I), English
translation from French by G. M. Temmer. North Holland, John Wiley \& Sons (1966).

\bibitem {muller}F. A. Muller, The equivalence myth of quantum mechanics; Part
I: Stud. Hist. Phil. Mod. Phys. \textbf{28}(1), 35--61 (1997); Part II: Stud.
Hist. Phil. Mod. Phys. \textbf{28}(2), 219--241 (1997).

\bibitem {Schiff}L. I. Schiff, \textit{Quantum Mechanics.} McGraw-Hill, New
York, 3d edition (1968).

\bibitem {schr}E. Schr\"{o}dinger, \"{U}ber das Verh\"{a}ltnis der
Heisenberg-Born-Jordanschen Quantenmechanik zu der meinen, Annalen der Physik
\textbf{79}, 734--756 (1926)

\bibitem {Song}D. Song, The P versus NP Problem in Quantum Physics,
arXiv:1402.6970 [physics.gen-ph]

\bibitem {Ville}V. Turunen, private communication.

\bibitem {sources}B. L. van der Waerden, ed. \textit{Sources of quantum
mechanics}. Courier Dover Publications (2007).

\bibitem {vanw}B. L. van der Waerden, From Matrix Mechanics and Wave Mechanics
to Unified Quantum Mechanics, Notices of the Amer. Math. Soc. \textbf{44}(3),
323--328 (1997), reprinted from \textit{The Physicist's Conception of Nature},
J. Mehra (ed.), 276--293, D. Reidel Publishing Company, Dordrecht, Holland (1973).

\bibitem {von}J. von Neumann, \textit{Mathematical foundations of quantum
mechanics}, Princeton University Press, Princeton (1955).

\bibitem {sh87}M. A. Shubin, \textit{Pseudodifferential Operators and Spectral
Theory}, Springer-Verlag, (1987) [original Russian edition in Nauka, Moskva (1978)].

\bibitem {Weyl}H. Weyl, Quantenmechanik und Gruppentheorie, Z. Physik
\textbf{46} (1927).
\end{thebibliography}
\end{document}